\documentclass[12pt,letter]{article}

\usepackage{graphicx, epsfig, color}
\def\eslt{\not\!\!{E_T}}
\def\to{\rightarrow}

\def\bi{\begin{itemize}}
\def\ei{\end{itemize}}

\def\sps1ap{SPS1a$^\prime$}
\def\c1p{C1$^\prime$}

\def\tb{\tilde b}

\def\tst{\tilde t}

\def\tg{\tilde g}

\def\tell{\tilde\ell}
\def\tq{\tilde q}
\def\tw{\widetilde W}
\def\tz{\widetilde Z}
\def\alt{\stackrel{<}{\sim}}
\def\agt{\stackrel{>}{\sim}}
\def\be{\begin{equation}}  
\def\ee{\end{equation}}  
\def\bea{\begin{eqnarray}}  
\def\eea{\end{eqnarray}}  
\def\beas{\begin{eqnarray*}}  
\def\eeas{\end{eqnarray*}}  
\newcommand\prd[3]{{\it Phys.\ Rev.\ }{\bf D #1} (#2) #3}

\newcommand\prl[3]{{\it Phys.\ Rev.\ Lett.\ }{\bf #1} (#2) #3}
\newcommand\plb[3]{{\it Phys.\ Lett.\ }{\bf B #1} (#2) #3}
\newcommand\jhep[3]{{\it J. High Energy Phys.\ }{\bf #1} (#2) #3}

\newcommand\npb[3]{{\it Nucl.\ Phys.\ }{\bf B #1} (#2) #3}

\newcommand{\hepph}[1]{hep-ph/#1}

\newcommand\ppnp[3]{{\it Prog.\ Part.\ Nucl.\ Phys.}{\bf  #1} (#2) #3}

\begin{document}
\begin{titlepage}

\vspace{0.5cm}
\begin{center}
{\Large \bf Implications of naturalness for \\
the heavy Higgs bosons of supersymmetry
}\\ 
\vspace{1.2cm} \renewcommand{\thefootnote}{\fnsymbol{footnote}}
{\large Kyu Jung Bae$^1$\footnote[1]{Email: bae@nhn.ou.edu },
Howard Baer$^1$\footnote[2]{Email: baer@nhn.ou.edu }, 
Vernon Barger$^2$\footnote[3]{Email: barger@pheno.wisc.edu },\\
Dan Mickelson$^1$\footnote[4]{Email: mickelso@nhn.ou.edu }
and Michael Savoy$^1$\footnote[5]{Email: savoy@nhn.ou.edu }
}\\ 
\vspace{1.2cm} \renewcommand{\thefootnote}{\arabic{footnote}}
{\it 
$^1$Dept. of Physics and Astronomy,
University of Oklahoma, Norman, OK 73019, USA \\
}
{\it 
$^2$Dept. of Physics,
University of Wisconsin, Madison, WI 53706, USA \\
}

\end{center}

\vspace{0.5cm}
\begin{abstract}
\noindent 
Recently, it has been argued that various measures of SUSY naturalness-- 
electroweak, Higgs mass and EENZ/BG-- 
when applied consistently concur with one another and make very specific predictions for 
natural supersymmetric spectra. 
Highly natural spectra are characterized by light higgsinos with mass not too far from 
$m_h$ and well-mixed but TeV-scale third generation squarks. 
We apply the unified naturalness measure to the case of heavy Higgs bosons $A$, $H$ and $H^\pm$. 
We find that their masses  are bounded from above by naturalness depending on $\tan\beta$: 
{\it e.g.} for 10\% fine-tuning and $\tan\beta\sim 10$, we expect $m_A\alt 2.5$ TeV whilst for 3\% 
fine-tuning and $\tan\beta$ as high as 50, then $m_A\alt 8$ TeV. 
Furthermore, the presence of light higgsinos seriously alters the heavy Higgs boson
branching ratios, thus diminishing prospects for usual searches into Standard Model (SM) 
final states, while new discovery possibilities arise due to the supersymmetric decay modes. 
The heavy SUSY decay modes tend to be $H,\ A,\ H^\pm\to W,\ Z,\ {\rm or}\ h+\eslt +{\rm soft\ tracks}$ 
so that single heavy Higgs production
is characterized by the presence of high $p_T$ $W$, $Z$ or $h$ bosons plus missing $E_T$.
These new heavy Higgs boson signatures seem to be challenging to extract from SM backgrounds.
\vspace*{0.8cm}

\end{abstract}

\end{titlepage}

\section{Introduction}
\label{sec:intro}

The recent discovery of a Standard Model like Higgs boson with mass $m_h=125.5\pm 0.5$ 
GeV\cite{atlas_h,cms_h} 
is in accord with predictions from supersymmetric models like the MSSM which require
$m_h\alt 135$ GeV\cite{mhiggs}. Such a large value of $m_h$ apparently requires TeV-scale top squarks
which are highly mixed, {\it i.e.} a large trilinear soft SUSY breaking parameter $A_t$\cite{h125}.
Coupling this result with recent SUSY search limits from LHC8\cite{atlas_susy,cms_susy} 
(which require $m_{\tg}\agt 1.3$ TeV for $m_{\tg}\ll m_{\tq}$ and $m_{\tg}\agt 1.8$ TeV for 
$m_{\tg}\sim m_{\tq}$) imply, within the context of gravity-mediated SUSY breaking models (SUGRA), 
a soft breaking scale characterized by a gravitino mass $m_{3/2}\agt 2$ TeV.
Indeed, a rather large SUSY breaking scale in gravity mediation models had been long
anticipated via a decoupling solution to the SUSY flavor, CP, proton decay and gravitino 
problems\cite{dine}.

In contrast, simple considerations of SUSY naturalness anticipate a SUSY breaking scale around
the weak scale typified by $m_Z\sim m_h\sim 100$ GeV. Thus, the Higgs mass and
sparticle mass limits combine to sharpen the ``Little Hierarchy''\cite{LH} typified by $m_h\ll m_{3/2}$.
The growing Little Hierarchy has prompted several authors to question whether the MSSM
is overly fine-tuned, and either flatly wrong\cite{shifman} or at least 
in need of additional features which sacrifice parsimony/minimality\cite{craig}.
Before rushing to such drastic conclusions, 
it is prudent to ascertain if all SUSY spectra are fine-tuned
or if some spectra are indeed natural. 

\subsection{Review of SUSY naturalness}

To proceed further one must adopt at least one of several quantitative naturalness 
measures which are available. 
We label these as
\begin{itemize}
\item the electroweak measure $\Delta_{EW}$\cite{ltr,sug,rns,perel,ccn},
\item the Higgs mass fine-tuning measure $\Delta_{HS}$\cite{fathiggs,kn} and
\item the traditional EENZ/BG measure $\Delta_{BG}$\cite{eenz,bg}.
\end{itemize}
Indeed, recently it has been shown that, if applied properly, then all three measures agree 
with one another\cite{dew} and predict a very specific SUSY spectra with just $\sim 10\%$ fine-tuning. 
If applied incorrectly-- by not properly combining {\it dependent} quantities 
contributing to $m_Z$ or $m_h$ one with another-- 
then overestimates\cite{comp} of fine-tuning can occur in $\Delta_{HS}$ and $\Delta_{BG}$, 
often by orders of magnitude.  

\subsubsection{$\Delta_{EW}$}

The electroweak measure $\Delta_{EW}$ requires that there be no large/unnatural cancellations
in deriving the value of $m_Z$ from the weak scale scalar potential:
\be
\frac{m_Z^2}{2} = \frac{(m_{H_d}^2+\Sigma_d^d)-(m_{H_u}^2+\Sigma_u^u)\tan^2\beta}{(\tan^2\beta -1)}
-\mu^2\simeq -m_{H_u}^2-\mu^2
\label{eq:mzs}
\ee
where $m_{H_u}^2$ and $m_{H_d}^2$ are the {\it weak scale} soft SUSY breaking Higgs masses, $\mu$ 
is the {\it supersymmetric} higgsino mass term and $\Sigma_u^u$ and $\Sigma_d^d$ contain
an assortment of loop corrections to the effective potential. The $\Delta_{EW}$ measure asks for
the largest contribution on the right-hand-side to be comparable to $m_Z^2/2$ so that no
unnatural fine-tunings are required to generate $m_Z=91.2$ GeV. The main requirement is then that
$|\mu |\sim m_Z$ and also that $m_{H_u}^2$ is driven radiatively to small, and not large, negative
values\cite{ltr,rns}. 
Also, the top squark contributions $\Sigma_u^u(\tst_{1,2})$ are minimized for TeV-scale highly
mixed top squarks, which also lift the Higgs mass to $m_h\sim 125$ GeV\cite{ltr}.

\subsubsection{$\Delta_{HS}$}

The Higgs mass fine-tuning measure $\Delta_{HS}$ asks that the radiative correction 
$\delta m_{H_u}^2$ to the Higgs mass
\be
m_h^2\simeq \mu^2+m_{H_u}^2(\Lambda )+\delta m_{H_u}^2
\label{eq:mhs}
\ee
be comparable to $m_h^2$. 
This contribution is usually written as $\delta m_{H_u}^2|_{rad}\sim -\frac{3f_t^2}{8\pi^2}(m_{Q_3}^2+m_{U_3}^2+A_t^2)\ln\left(\Lambda^2/m_{SUSY}^2 \right) $ which is used to claim that third generation squarks
$m_{\tst_{1,2},\tb_1}$ be approximately less than 500 GeV and $A_t$ be small for natural SUSY. 
However, several
approximations are necessary to derive this result, the worst of which is to neglect that
the value of $m_{H_u}^2$ itself contributes to $\delta m_{H_u}^2$. By combining dependent
contributions, then instead one requires that the two terms on the RHS of 
\be
m_h^2 = \mu^2+\left( m_{H_u}^2(\Lambda) +\delta m_{H_u}^2\right)
\label{eq:mh2}
\ee 
be comparable to $m_h^2$.\footnote{It is sometimes claimed that by using this method, then the SM 
would not be fine-tuned for large cutoff scales $\Lambda\gg 1$ TeV. However-- in contrast to the SM-- 
for the SUSY case, EW symmetry is not even broken at tree level in models where the soft terms
arise from hidden sector SUSY breaking. Further discussion of the differences is included in Ref's 
\cite{comp,dew}.} The recombination in Eq. \ref{eq:mh2} leads back to
the EW measure since $m_{H_u}^2(\Lambda )+\delta m_{H_u}^2 =m_{H_u}^2(weak)$.

\subsubsection{$\Delta_{BG}$}

The EENZ/BG measure\cite{eenz,bg} (hereafter denoted simply by BG) is given by
\be
\Delta_{BG}\equiv max_i\left[ c_i\right]\ \ {\rm where}\ \ c_i=\left|\frac{\partial\ln m_Z^2}{\partial\ln p_i}\right|
=\left|\frac{p_i}{m_Z^2}\frac{\partial m_Z^2}{\partial p_i}\right|
\label{eq:DBG}
\ee
where the $p_i$ constitute the fundamental parameters of the model.
Thus, $\Delta_{BG}$ measures the fractional change in $m_Z^2$ due to fractional variation in 
(high scale) parameters $p_i$. 
The $c_i$ are known as {\it sensitivity co-efficients}\cite{bg}.
For the pMSSM (MSSM defined only at the weak scale), then explicit evaluation gives 
$\Delta_{BG}\simeq \Delta_{EW}$.
For models defined in terms of high scale parameters, the BG measure can be evaluated by 
expanding the terms on the RHS of Eq. \ref{eq:mzs}
using semi-analytic RG solutions in terms of fundamental high scale parameters\cite{munoz}:
for $\tan\beta =10$ and taking $\Lambda =m_{GUT}$, then one finds\cite{abe,martin}
\be
m_Z^2 \simeq  -2.18\mu^2 + 3.84 M_3^2-0.65 M_3A_t-1.27 m_{H_u}^2 -0.053 m_{H_d}^2
+0.73 m_{Q_3}^2+0.57 m_{U_3}^2 +\cdots
\label{eq:mzsHS}
\ee
The BG measure picks off the co-efficients of the various terms and recales by the soft term
squared over the $Z$-mass squared: {\it e.g.} $c_{m_{Q_3}^2}=0.73\cdot (m_{Q_3}^2/m_Z^2)$. 
If one allows $m_{Q_3}\sim 3$ TeV (in accord with 
requirements from the measured value of $m_h$) then one obtains $c_{m_{Q_3}^2}\sim 800$
and so $\Delta_{BG}\ge 800$. In this case, SUSY would be electroweak fine-tuned to about 0.1\%. 

The problem with most applications of the BG measure  is that in any sensible model of SUSY breaking, 
the high scale SUSY parameters are not independent. For instance , in gravity-mediation, then 
for any given hidden sector, the SUSY soft breaking terms are all calculated as numerical
co-efficients times the gravitino mass\cite{sw,kl,Brignole:1993dj}: {\it e.g.} $M_3(\Lambda )=a_{M_3} m_{3/2}$, 
$A_t=a_{A_t}m_{3/2}$, $m_{Q_3}^2=a_{Q_3}m_{3/2}^2$ {\it etc.} where the $a_i$ are just numbers. 
(For example, in string theory with dilaton-dominated SUSY breaking\cite{kl,Brignole:1993dj}, 
then we expect $m_0^2=m_{3/2}^2$ with $m_{1/2}=-A_0=\sqrt{3}m_{3/2}$).
The reason one scans multiple SUSY model soft term parameters is to account for a 
wide variety of possible hidden sectors. 
But this doesn't mean each soft term is independent from the others.
By writing the soft terms in Eq. \ref{eq:mzsHS} as suitable multiples of $m_{3/2}^2$, 
then large positive and negative contributions can be combined/cancelled 
and one arrives at the simpler expression\cite{dew}:
\be
m_Z^2=-2.18\mu^2(\Lambda) +a\cdot m_{3/2}^2 .
\label{eq:mzssugra}
\ee
The value of $a$ is just some number which is the sum of all the coefficients of the terms 
$\propto m_{3/2}^2$.\footnote{If $\mu$ is also computed as $\mu =a_\mu m_{3/2}$ as in the Giudice-Masiero
mechanism\cite{Giudice:1988yz}, 
then $m_Z^2=const.\times m_{3/2}^2$ and $\Delta_{BG}\equiv 1$ and there would be no 
fine-tuning\cite{dimitri}.
In other solutions of the SUSY $\mu$-problem, such as Kim-Nilles\cite{Kim:1983dt}, 
then $\mu$ is instead related to the
Peccei-Quinn breaking scale and is expected to be independent. In the former case, then the 
responsibility is to find a suitable hidden sector which would actually 
generate $m_Z^2$ at its measured value. We are aware of no such models which even come close to that.}
Using the BG measure applied to Eq. \ref{eq:mzssugra}, then it is found that naturalness requires 
$\mu^2\sim m_Z^2$ and also that $am_{3/2}^2\sim m_Z^2$. The first requirement is the same as in
$\Delta_{EW}$. The second requirement is fulfilled {\it either} by $m_{3/2}\sim m_Z$\cite{bg} 
(which seems unlikely in light of LHC Higgs mass measurement and sparticle mass bounds) {\it or}
that $m_{3/2}$ is large but the co-efficient $a$ is small\cite{dew}: {\it i.e.} there are large cancellations
in Eq. \ref{eq:mzsHS}. Since $\mu (\Lambda )\simeq \mu (weak)$, then also 
$am_{3/2}^2\simeq m_{H_u}^2(weak)$ and so a low value of $\Delta_{BG}$ also requires a low value
of $m_{H_u}^2$: {\it i.e.} $m_{H_u}^2$ is driven radiatively to small negative values.
This latter situation is known as {\it radiatively-driven natural supersymmetry}, or RNS.

\subsection{Naturalness and heavy SUSY Higgs bosons}

The natural SUSY spectra is typified by a spectra of low-lying Higgsinos $\tw_1^\pm$, $\tz_{1,2}$
with mass $\sim 100-300$ GeV, the closer to $m_Z$ the better, along with TeV-scale but highly mixed
top-squarks $\tst_{1,2}$.\cite{ltr,rns} 
The gluino mass can range between current LHC8 limits and about 4 TeV, and may well lie
beyond LHC14 reach\cite{rns@lhc}. First/second generation matter scalars may well lie in the $5-30$
TeV range, thus supplying at least a partial decoupling solution to the SUSY flavor, CP, proton decay
and gravitino problem\footnote{Since $m_{\tq,\tell}\sim m_{3/2}$, then we would expect
$m_{3/2}$ also at the $5-30$ TeV level.}. 
In addition, it should be clear from Eq. \ref{eq:mzs} that $m_{H_d}^2/\tan^2\beta\sim m_Z^2$ 
(a point mentioned previously in Ref. \cite{bbht}). 
For $m_{H_d}$ large, then one expects $m_A\sim m_{H_d}$. 
Requiring the term containing $m_{H_d}^2$ in Eq. \ref{eq:mzs} to be
comparable to $m_Z^2/2$ or $\mu^2$ then implies
\be
m_A\sim \left|m_{H_d}^2\right|^{1\over 2} \alt |\mu | \tan\beta \ .
\label{eq:mA}
\ee
Thus, for $|\mu|<300$ GeV, we would expect for $\tan\beta =10$ that $m_A\alt 3$ TeV. 
But for $\tan\beta$ as high as 50, we expect $m_A\alt 15$ TeV without becoming too unnatural.

In this paper, we explore the implications of SUSY naturalness for the heavy Higgs bosons of the MSSM:
$A$, $H$ and $H^\pm$. This topic has also been addressed in the recent paper \cite{Katz:2014mba}.
In Ref. \cite{Katz:2014mba}, using several different naturalness measures along with a low mediation scale
$\Lambda\sim 10-100$ TeV and hard SUSY breaking contributions to the scalar potential,
the authors conclude that heavy Higgs bosons should lie around the 1 TeV scale, and that 
since the heavy Higgs bosons are less susceptible to having hidden decay modes, their search
should be an important component of the search for natural SUSY.

In this paper, we will arrive at quite different conclusions. In Sec. \ref{sec:mass}, 
using the unified naturalness criteria, as embodied in $\Delta_{EW}$, we will find that
SUSY models which are valid all the way up to $\Lambda=m_{GUT}\simeq 2\times 10^{16}$ GeV
can be found with fine-tuning at the $\Delta_{EW}\sim 7-30$ level, corresponding to mild
fine-tunings of just 3-15\%. In this case, then as suggested in Eq. \ref{eq:mA}, we find
that natural SUSY models exist with $m_A\alt 5$ TeV for $\tan\beta \alt 15$ while
$m_A\alt 8$ TeV for $\tan\beta$ values ranging as high as $50-60$. 
While the region $m_A\alt$ 1 TeV should be accesible to LHC heavy Higgs searches, 
the bulk of this mass range is well beyond any projected LHC reach.
In Sec. \ref{sec:lhc}, we evaluate the heavy Higgs $A$, $H$ and $H^\pm$ branching fractions
as a function of mass for a benchmark case with radiatively-driven naturalness. Since 
for naturalness $\mu\sim 100-300$ GeV, then the heavy Higgs decays to higgsino pairs is almost always open.
Since the higgsinos are essentially invisible at LHC, these modes lead to quasi-invisible decays.
Further, since the heavy Higgs coupling to the -ino sector 
(here, -ino collectively refers to both charginos and neutralinos)
is a product of gaugino times higgsino
components, then once kinematically accesssible, the heavy Higgs tend to decay dominantly into
gaugino plus higgsino states. Such large branching fractions reduce the heavy Higgs branching fractions into 
SM modes, making standard heavy Higgs searches more difficult. Alternatively, 
since the gauginos tend to decay to gauge/Higgs bosons $W$, $Z$ or $h$ plus higgsinos, 
then the qualitatively new decay modes arise: $A$, $H$, $H^\pm\to$ $W$, $Z$ or $h$ plus 
missing $E_T$ ($\eslt$). These new decay modes-- which are quite different than those expected
in non-natural SUSY models with a bino-like LSP-- offer new avenues for heavy Higgs searches at
LHC.

\section{Mass bounds from naturalness}
\label{sec:mass}

A simple mass bound from naturalness on heavy Higgs bosons can be directly read off from 
Eq. \ref{eq:mzs}. The contribution to $\Delta_{EW}$ from the $m_{H_d}^2$ term is given 
by
\be
C_{H_d}= m_{H_d}^2/(\tan^2\beta -1)/ (m_Z^2/2) .
\ee
Also the tree level value of $m_A$ is given by
\be
m_A^2=m_{H_u}^2+m_{H_d}^2+2\mu^2\simeq m_{H_d}^2-m_{H_u}^2\sim m_{H_d}^2
\ee
where the first partial equality holds when $\mu^2\sim -m_{H_u}^2$ and the second
arises when $m_{H_d}^2\gg -m_{H_u}^2$.
Combining these equations, then one expects roughly that
\be
m_A\alt m_Z\ \tan\beta\    \Delta_{EW}^{1/2}(max) 
\ee
where $\Delta_{EW}(max)$ is the maximal fine-tuning one is willing to tolerate.
For $\Delta_{EW}^{-1}=10\%$ fine-tuning with $\tan\beta =10$, then one 
expects $m_A\alt 3$ TeV.

However, this simple argument is not the whole story since an assortment of radiative corrections
are included in Eq. \ref{eq:mzs}. 
In particular, the radiative corrections $\Sigma_u^u(\tst_{1,2})$ and $\Sigma_u^u(\tb_{1,2})$ 
(complete expressions are provided in the appendix of Ref. \cite{rns}) can become 
large and are highly $\tan\beta$ dependent.

To evaluate the range of $m_A$ expected by naturalness, 
we will generate SUSY spectra using Isajet\cite{isajet,isasugra} in the 
2-parameter non-universal Higgs model\cite{nuhm2} (NUHM2) which allows for very low values of
$\Delta_{EW}<10$ (numerous other constrained models are evaluated in Ref. \cite{dew} and always
give much higher EW fine-tuning).
The parameter space is given by
\be
m_0,\ m_{1/2},\ A_0,\ \tan\beta,\ \mu,\ m_A,\  \qquad {\rm (NUHM2)}. 
\label{eq:nuhm2}
\ee
The NUHM2 spectra and parameter spread versus $\Delta_{EW}$ were evaluated in Ref. \cite{rns}
but with the range of $m_A$ restricted to $<1.5$ TeV. Here, we improve this scan by 
including a much large range of $m_A$:
\bea
m_0 &:& \ 0-20\ {\rm TeV}, \nonumber\\
m_{1/2} &:& \  0.3-2\ {\rm TeV},\nonumber\\
-3 &<& A_0/m_0 \ <3,\nonumber\\
\mu &:& \ 0.1-1.5\ {\rm TeV}, \label{eq:param}\\
m_A &:& \ 0.15-20\ {\rm TeV},\nonumber\\
\tan\beta &:& 3-60 . \nonumber
\eea
We require of our solutions that:
\bi
 \item electroweak symmetry be radiatively broken (REWSB),
 \item the neutralino $\tz_1$ is the lightest MSSM particle,
 \item the light chargino mass obeys the model
independent LEP2 limit, $m_{\tw_1}>103.5$~GeV\cite{lep2ino},
\item LHC search bounds on $m_{\tg}$ and $m_{\tq}$ are respected,
\item $m_h=125.5\pm 2.5$~GeV.
\ei

\begin{figure}[tbp]
\includegraphics[height=0.5\textheight]{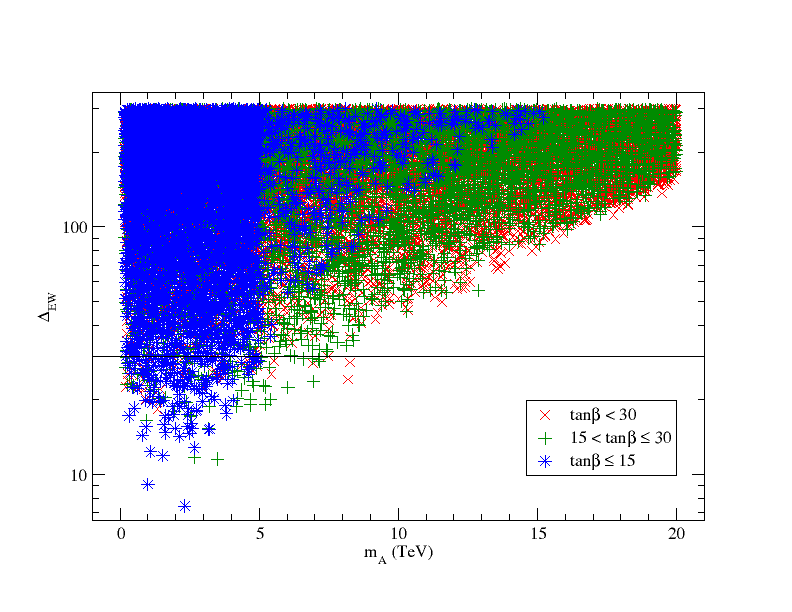}
\caption{Plot of $\Delta_{EW}$ versus $m_A$ from a scan over NUHM2 parameter space.
\label{fig:scan}}
\end{figure}

The results of our scan are shown in Fig. \ref{fig:scan} where we plot $\Delta_{EW}$ vs. $m_A$.
The dots are color-coded according to low, intermediate and high $\tan\beta$ values.
From the plot, we see first that there is indeed an upper bound to $m_A$ given by naturalness.
In fact, for $\tan\beta <15$ and $\Delta_{EW}<10$, then indeed $m_A\alt 3$ TeV as 
suggested by the simple arguments above. For $\tan\beta >15$, we do not generate any
solutions with $\Delta_{EW}<10$. For $\Delta_{EW}<30$ (dotted horizontal line), then
we have $m_A\alt 5$ TeV for $\tan\beta <15$, and $m_A\alt 7$ (8) TeV for
$\tan\beta <30$ (60). While these values provide upper bounds on $m_A$ from naturalness, 
we note that $m_A$ values as low as 150-200 GeV can also be found. Since LHC14 searches for
heavy Higgs are roughly sensitive to $m_A\alt 1$ TeV values\cite{lhchiggsplane}, 
then we conclude that LHC14 searches will be able to probe a portion of natural 
SUSY parameter space, but perhaps the 
bulk of parameter space can easily lie well  beyond Atlas/CMS search capabilities.

\begin{figure}[tbp]
\includegraphics[height=0.5\textheight]{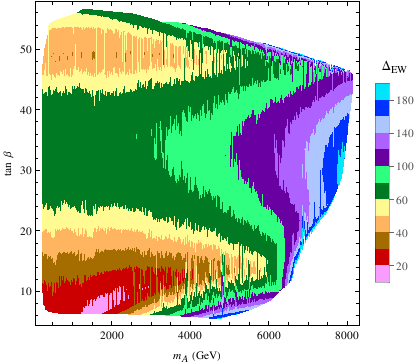}
\caption{Regions of SUSY naturalness $\Delta_{EW}$ in the $m_A$ vs. $\tan\beta$ plane
for the RNS benchmark point Eq. \ref{eq:BM}.
\label{fig:plane}}
\end{figure}

To gain more perspective on fine-tuning and how it depends on $m_A$ and $\tan\beta$, we
next adopt a proposed RNS benchmark point from Ref. \cite{ilcbm8}. 
This point has NUHM2 parameters given by
\be
m_0=5\ {\rm TeV},\ m_{1/2}=0.7\ {\rm  TeV},\ A_0=-8.3\ {\rm TeV},\ \tan\beta =10,\ {\rm with}\ \mu =110\ {\rm GeV}\ 
{\rm and}\ m_A=1\ {\rm TeV}.
\label{eq:BM}
\ee
The value of $\Delta_{EW}$ is found to be 13.8 .
Here, we adopt this benchmark point, but now allow  $m_A$ and $\tan\beta$ as free parameters
and plot color-coded ranges of $\Delta_{EW}$ in the $m_A$ vs. $\tan\beta$ plane, 
as shown in Fig. \ref{fig:plane}.

From Fig. \ref{fig:plane}, we see that indeed the region with lowest $\Delta_{EW}$
occurs around $m_A\sim 1.2-2.5$ TeV with $\tan\beta \alt 10$. 
The yellow colored regions have $\Delta_{EW}<50$. For these values, we find a more expansive region
with $m_A\alt 6$ TeV and $\tan\beta \alt 20$. However, a second region with low $\Delta_{EW}<50$
opens up at high $\tan\beta\sim 48-52$ with $m_A\alt 4$ TeV. The intermediate $\tan\beta\sim 20-45$ 
region has greater fine-tuning, where the maximal contributions to $\Delta_{EW}$ we find 
arise from the radiative corrections $\Sigma_u^u(\tb_2 )$.

\section{Implications for heavy Higgs discovery at LHC} 
\label{sec:lhc}

In many studies of the prospects for heavy Higgs boson discovery at the LHC, 
it is assumed that the Standard Model decay modes of $A$, $H$ and $H^\pm$ 
are dominant. The prospects for discovery are usually presented in the 
$m_A$ vs. $\tan\beta$ plane. At NLO in QCD, then the gluon fusion
reactions $gg\to A,\ H$ are usually dominant out to $m_{A,H}\alt 1$ TeV
while for higher masses then vector boson fusion (VBF) dominates\cite{Spira:1995rr}.
The main discovery mode for $gg\to A,\ H$ is then the $A,\ H\to \tau^+\tau^-$
mode where the ditau mass can be reconstructed. Current search limits from
Atlas and CMS exclude $m_A\alt 0.8$ TeV for $\tan\beta$ as high as 50. 
For lower $\tan\beta$ values, the mass bounds are very much weaker\cite{Flechl:2013wza}
({\it e.g.} for $\tan\beta =10$, then $m_A>400$ GeV).
Production of heavy Higgs bosons in association with $b$-jets may aid the search\cite{Kao:2007qw}.
In addition, the rarer decays into dimuons may also be possible\cite{Kao:1995gx,Barger:1997pp}, 
and recently dimuon signatures in association with $b$-jets have been explored\cite{Dawson:2002cs,Baer:2011ua}.

The importance of heavy Higgs decay into SUSY modes was explored long ago\cite{Baer:1992kd}
for the case where the LSP was usually taken to be a bino.
If SUSY decay modes of $H$ or $A$ are open, then the SM branching fractions diminish
while the new SUSY modes offer novel detection strategies\cite{Baer:1994fx}.

\subsection{Heavy Higgs branching fractions in natural SUSY}
\label{ssec:BFs}

The unique feature of SUSY models with radiatively-driven naturalness is the presence of light
higgsino states with mass $\sim 100-300$ GeV, the closer to $m_Z$ the better. 
This fact means that for most of the mass range of $m_{A,H}$, then SUSY decay modes should be open.
Furthermore, the higgsino-like LSP implies that the SUSY decay modes will generally be quite
different than in earlier models where a bino-like LSP was considered.

\begin{figure}[tbp]
\includegraphics[height=0.5\textheight]{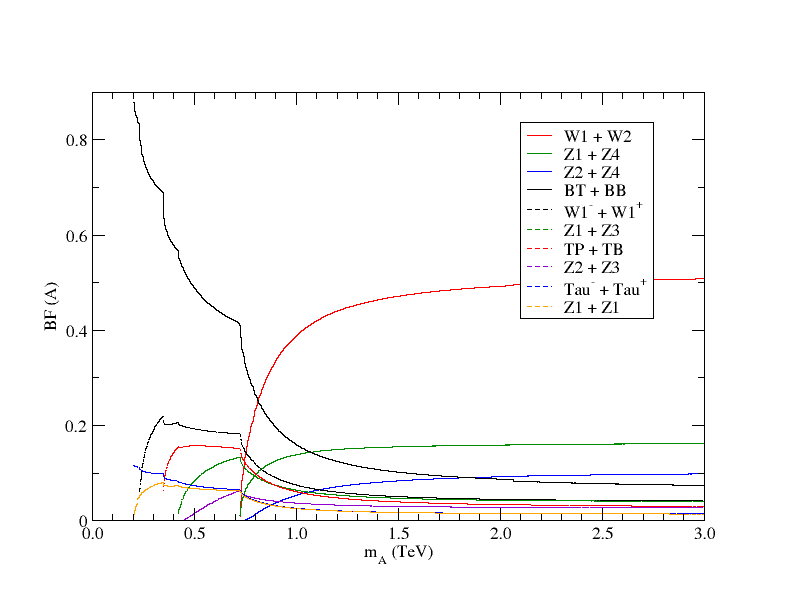}
\caption{Branching fraction of $A$ vs. $m_A$ for the RNS benchmark point Eq. \ref{eq:BM} but with variable $m_A$.
\label{fig:BFA}}
\end{figure}

In Fig. \ref{fig:BFA}, we show the branching fraction as calculated by Isajet\cite{isajet} 
of the pseudoscalar $A$ boson versus
$m_A$ for the RNS benchmark point from Sec. \ref{sec:mass}, but now with $m_A$ taken as variable, with
$\tan\beta =10$. At low $m_A\sim 200$ GeV, then SUSY decay modes are kinematically closed
and $A\to b\bar{b}$ at $\sim 85\%$ as is typical when SM decay modes are considered and the $t\bar{t}$
mode is closed. As $m_A$ increases beyond $200$ GeV, then already the $A\to higgsino\ pairs$
opens up, and the SM branching fractions diminish. For $m_A\agt 700$ GeV, then the mixed higgsino/wino mode
$A\to\tw_1\tw_2$ turns on and rapidly dominates the branching fraction. 
This is because the SUSY Higgs coupling to -inos involves a product of 
gaugino component of one -ino times the higgsino components of the other 
-ino\footnote{See p. 178-179 of \cite{wss}.} and in this case $\tw_1$ is 
higgsino-like and $\tw_2$ is wino-like.
For $m_A\agt 1$ TeV, 
this decay mode is typically at the $\sim 50\%$ level. For $m_A\agt 1$ TeV, then also the decays
$A\to \tz_1\tz_4$ and $\tz_2\tz_4$ are important. 
For TeV-scale values of $m_A$, the SM decay mode $A\to b\bar{b}$ drops to below the 10\% level 
while $A\to\tau\bar{\tau}$ has dropped to the percent level.
In this case, then the search for heavy Higgs bosons utilizing SM decay modes
will be much more difficult.

\begin{figure}[tbp]
\includegraphics[height=0.5\textheight]{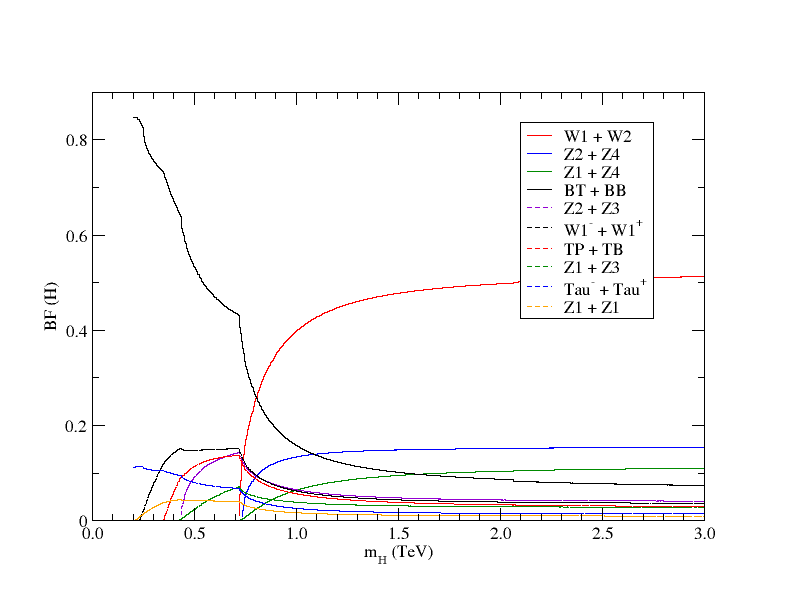}
\caption{Branching fraction of $H$ vs. $m_H$ for the RNS benchmark point Eq. \ref{eq:BM} but with variable $m_A$.
\label{fig:BFH}}
\end{figure}

In Fig. \ref{fig:BFH}, we show the branching fractions of the heavy scalar Higgs $H$ versus
$m_H$ for the same RNS benchmark point. The overall behavior is similar to the case of the
pseudoscalar $A$: at low values of $m_H$, then the SM decay modes are dominant, but once
$m_H$ is heavy enough, the supersymmetric decay modes quickly open up and dominate the
branching fractions. At large $m_H$, then the $H\to \tw_1\tw_2$, $\tz_2\tz_4$ and $\tz_1\tz_4$
decay modes are dominant.

\begin{figure}[tbp]
\includegraphics[height=0.5\textheight]{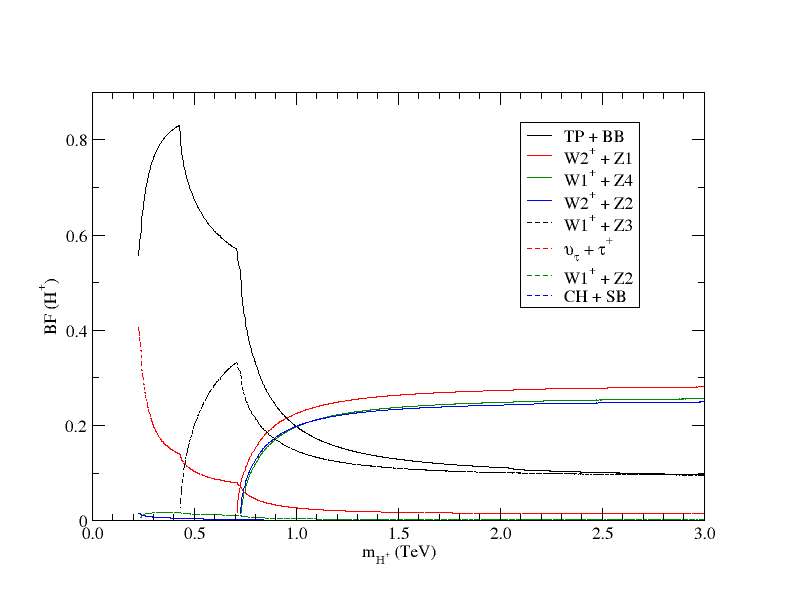}
\caption{Branching fraction of $H^+$ vs. $m_{H^+}$ for the RNS benchmark point Eq. \ref{eq:BM} but with variable $m_A$.
\label{fig:BFHC}}
\end{figure}

In Fig. \ref{fig:BFHC} we show the branching fractions of $H^+$ versus $m_{H^+}$ for the 
same RNS benchmark point. In this case, at low values of $m_{H^+}$, then $H^+\to t\bar{b}$ 
is dominant followed by $H^+\to \tau^+\nu_\tau$. As $m_{H^+}$ increases, then 
$H^+\to \tw_1^+\tz_3$ turns on and later also $\tw_2^+\tz_1$, $\tw_1^+\tz_4$ and $\tw_2^+\tz_2$
all turn on. 
At $m_{H^+}\agt 1$ TeV, these latter decays into gaugino/higgsino final states dominate. 
Such non-standard decay modes will make searches for charged Higgs bosons more
difficult than otherwise expected\cite{Roy:2005yu}.

\subsection{New SUSY Higgs signatures at LHC}

\subsubsection{$H$, $A\to W+\eslt$}

We have seen that for $m_{A,H}\agt 1$ TeV, then the dominant branching fraction is
$H, A\to \tw_1^\pm\tw_2^\mp$. Since the $\tw_1$ is higgsino-like, it tends to have only a small mass
gap with the LSP: $m_{\tw_1}-m_{\tz_1}\sim 10-20$ GeV. In this case, the visible energy from
$\tw_1\to f\bar{f}'\tz_1$ decay (where $f$ denotes SM fermions) is quite soft-- 
most of the energy goes into making up the $\tz_1$ rest mass--
and so the higgsinos are only quasi-visible. On the other hand, the branching fractions for
$\tw_2$ decay in the RNS model have been plotted out in Ref. \cite{rns@lhc} and found to be:
$\tw_2\to \tw_1 Z$, $\tz_1 W$,  $\tz_2 W$ each at about 30\% with $\tz_3 W$ accounting for the remainder.
Thus, we expect $s$-channel $H$ and $A$ production to give rise  to 
\be
gg\to H,\ A\to W+\eslt \to \ell^\pm +\eslt
\ee
which is a rather unique signature for heavy Higgs boson production. 

\begin{figure}[tbp]
\includegraphics[height=0.5\textheight]{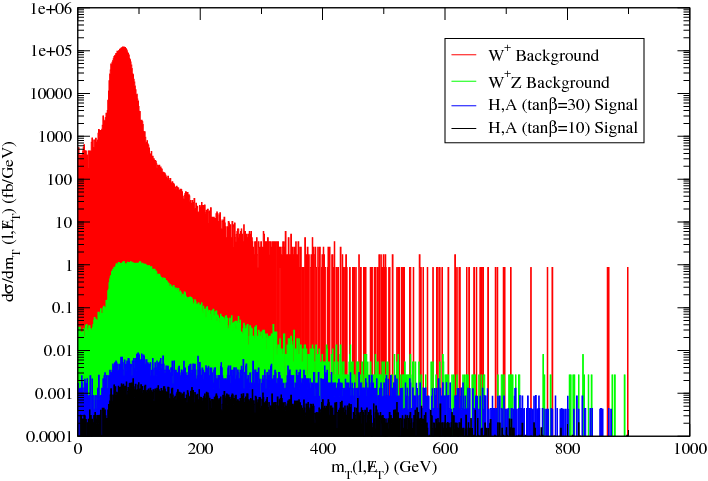}
\caption{Transverse mass distribution for $e^+$ plus $\eslt$ events at LHC14 from
$W^+$, $W^+Z$ and $A,\ H$ production from benchmark point Eq. \ref{eq:BM}.}
\label{fig:mT}
\end{figure}

The dominant backgrounds come from direct $W$ production followed by $W\to\ell\nu_{\ell}$ decay
and also $WZ$ production followed by $Z\to\nu\bar{\nu}$ and $W\to \ell\nu_{\ell}$.
In Fig.~\ref{fig:mT}, we plot the $e^++\eslt$ transverse mass distribution from the 
signal using the RNS benchmark point with $m_A=1$ TeV along with SM backgrounds.
The signal from $A,\ H$ production with $m_{A,H}\sim 1$ TeV and $\tan\beta =10$ 
and 30 is well below background.

\subsection{$H$, $A\to Z+\eslt$}

\begin{figure}[tbp]
\includegraphics[height=0.5\textheight]{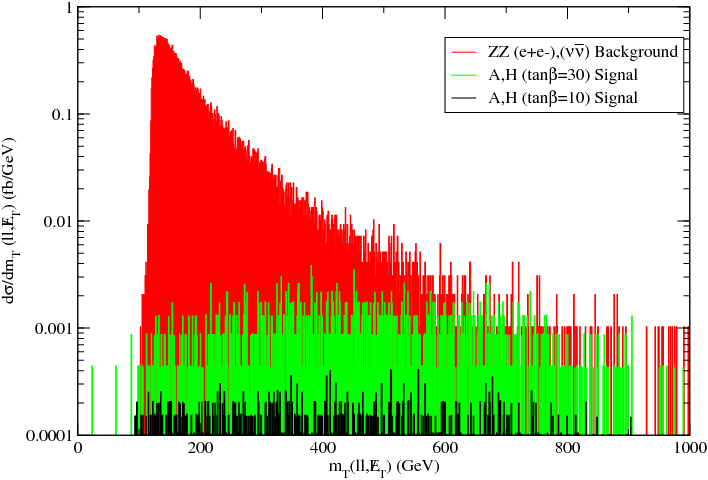}
\caption{Dilepton cluster transverse mass distribution for $e^+e^-$ plus $\eslt$ events at LHC14 from
$ZZ$ and $A,\ H$ production from benchmark point Eq. \ref{eq:BM}.}
\label{fig:mcT}
\end{figure}

As mentioned above, $\tw_2\to \tw_1 Z$ at about 30-35\% in radiatively-driven natural SUSY. 
Thus, an alternative signature comes from 
\be
gg\to H,\ A\to Z+\eslt \to \ell^+\ell^- +\eslt .
\ee
The background to this process comes from $ZZ$ production where one $Z\to\nu\bar{\nu}$
whilst the other goes as $Z\to\ell^+\ell^-$. 
In Fig. \ref{fig:mcT} we plot the distribution in cluster transverse mass\cite{vb} $m_T(\ell^+\ell^-,\eslt )$
from heavy Higgs $H$, $A$ production followed by their decays to $Z(\to\ell^+\ell^-)+\eslt$ from the
RNS benchmark point for $m_A=1$ TeV along with $ZZ$ background. 
Here we see that signal from $A,\ H$ production with $m_{A,H}\sim 1$ TeV lies well below
the diboson background for $\tan\beta =10$. If we increase $\tan\beta$ to 30, then signal and BG become comparable at very large $m_T(\ell^+\ell^-,\eslt)$ although in this range the event rate is quite limited.

\subsection{$H$, $A\to h+\eslt$}

A third possible signature consists of $A,\ H\to \tz_{1,2}\tz_{3,4}$ where 
$\tz_{3,4}\to \tz_{1,2} h$ resulting in a $(h\to b\bar{b})+\eslt$ signature.
We expect such a signal to lie well below backgrounds from $Zh$ and $ZZ$ production.

\section{Conclusions:} 
\label{sec:conclude}

In this paper we have examined the implications of SUSY naturalness for the heavy Higgs boson sector.
We use the $\Delta_{EW}$ measure of naturalness, although we show that-- properly applied--
the Higgs mass fine-tuning and also the EENZ/BG fine-tuning would give similar results
since 
\be
\Delta_{HS}\simeq \Delta_{BG}\simeq \Delta_{EW}
\ee
so long as {\it dependent} terms are properly combined before evaluating naturalness.

Using the $\Delta_{EW}$ measure, then we find upper bounds on the heavy Higgs masses:
for 10\% fine-tuning and $\tan\beta\sim 10$, we expect $m_A\alt 2.5$ TeV whilst for 3\% 
fine-tuning and $\tan\beta$ as high as 50, then $m_A\alt 8$ TeV.
These values are considerably larger than the range depicted in Ref. \cite{Katz:2014mba} and much of this range
likely lies beyond LHC14 reach. This means LHC14 will be able to probe only a portion--
but by no means all-- of natural SUSY parameter space via heavy Higgs boson searches.

Furthermore, since light higgsino states $\tw_1^\pm$ and $\tz_{1,2}$ are expected to have
mass $\sim 100-300$ GeV (the closer to $m_Z$ the more natural), then almost always there will
be supersymmetric decays modes open to the heavy SUSY Higgs states.
We evaluated these branching fractions and find that they can in fact be the dominant
decay modes, especially if $m_{A,H}>m_{\tw_1}+m_{\tw_2}$, in which case this decay mode tends to dominate.
The supersymmetric decay modes diminish the SM decay modes of $H$, $A$ and $H^\pm$ making
standard search techniques more difficult for a specified heavy Higgs mass.
However, qualitatively new heavy Higgs search modes appear thanks to the supersymmetric decay modes.
Foremost among these are the decays $H,\ A\to\tw_1^\pm\tw_2^\mp$ which results in 
final states characterized by $W$, $Z$ or $h$ plus $\eslt$. 
These new signatures seem to be rather challenging to extract from SM backgrounds which occur
at much higher rates.
It may well be that forward $b$-jet tagging in $bg\to bA$ or $bH$ production 
or $gb\to tH^+$ production followed by $A,\ H, H^\pm\to SUSY$ decays could 
ameliorate the situation. 

\section*{Acknowledgments}

We thank X. Tata for discussions.
This work was supported in part by the US Department of Energy, Office of High
Energy Physics.

%

%
\end{document}